\documentstyle[12pt,lnfprep,epsfig]{article}

\newcommand{\Header}{
  \begin{tabular}{rl}
  \hspace{-.4cm}\includegraphics{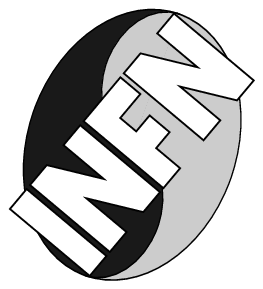} &
    \renewcommand{\arraystretch}{0.5}
    \begin{tabular}{r}
      {\hspace{1cm}~\LARGE\sffamily LABORATORI~ NAZIONALI~ DI~ FRASCATI}\\
      \\
      {\Large\sffamily SIS-Pubblicazioni}\\
    \end{tabular}
    \renewcommand{\arraystretch}{1}
  \end{tabular}
  \vskip 1cm
  \begin{flushright}
  \renewcommand{\arraystretch}{0.5}
    \begin{tabular}{r}
      {\underline{LNF-99/009(P)}}\\      
      {\small 19 Maggio 1999}     \\     
      \\
      {\small INFNNA-IV-99/12}\\
      {\small UWThPh-1999-19}\\
    \end{tabular}
  \end{flushright}
  \renewcommand{\arraystretch}{1}
  \vskip 1 cm
  }

\newcommand{\czdot}{\! \cdot \!}
\newcommand{\beq}{\begin{equation}}
\newcommand{\eeq}{\end{equation}}
\newcommand{\beqa}{\begin{eqnarray}}
\newcommand{\eeqa}{\end{eqnarray}}
\newcommand{\beqan}{\begin{eqnarray*}}
\newcommand{\eeqan}{\end{eqnarray*}}
\newcommand{\ba}{\begin{array}}
\newcommand{\ea}{\end{array}}
\newcommand{\ben}{\begin{enumerate}}
\newcommand{\een}{\end{enumerate}}
\newcommand{\bfl}{\begin{flushleft}}
\newcommand{\efl}{\end{flushleft}}
\newcommand{\btab}{\begin{tabular}}
\newcommand{\etab}{\end{tabular}}
\newcommand{\bit}{\begin{itemize}}
\newcommand{\eit}{\end{itemize}}
\newcommand{\bdes}{\begin{description}}
\newcommand{\edes}{\end{description}}
\newcommand{\bdm}{\begin{displaymath}}
\newcommand{\edm}{\end{displaymath}}
\newcommand{\nl}{\nonumber \\}

\newcommand{\ra}{\rightarrow}

\newcommand{\cL}{{\cal L}}

\begin{document}
\begin{titlepage}
\title{ 
  \Header
  {\large \bf $\eta \ra \pi^+\pi^- \pi^0 \gamma$ in Chiral 
               Perturbation Theory*}
\author{
  G. D'Ambrosio$^1$, G. Ecker$^2$, G. Isidori$^3$ and
  H. Neufeld$^2$ \\
  \\
 {\it ${}^{1)}$ INFN, Sezione di Napoli}\\
 {\it Dipartimento di Scienze Fisiche, Universit\`a di Napoli
     I--80126 Napoli, Italy} \\
 {\it ${}^{2)}$ Institut f\"ur Theoretische Physik, Universit\"at Wien}\\
 {\it Boltzmanngasse 5, A--1090 Wien, Austria} \\
 {\it ${}^{3)}$ INFN, Laboratori Nazionali di Frascati,
      P.O. Box 13, I--00044 Frascati, Italy }
}}
\maketitle

\baselineskip=14pt

\vskip 1.5 true cm
\begin{abstract}
We analyse the radiative decay $\eta \to \pi^+\pi^- \pi^0\gamma$  
in the low--energy expansion of the Standard Model.
We employ the notion of ``generalized bremsstrahlung'' to
take full advantage of the theoretical and experimental information on
the corresponding non-radiative $\eta \to 3 \pi$ decay. The direct
emission amplitude of ${\cal O}(p^4)$ is due to one-loop diagrams with
intermediate pions (isospin violating) and kaons (isospin conserving).
The isospin conserving contributions to direct emission, including
vector meson exchange appearing at ${\cal O}(p^6)$, are suppressed.
\end{abstract}

\vspace*{\stretch{2}}
\begin{flushleft}
\end{flushleft}
\begin{center}
Submitted to Physics Letters B
 \vskip 1cm
{\small * Work supported in part
by TMR, EC--Contract No. ERBFMRX--CT980169 (EURODA$\Phi$NE)}
\end{center}
\end{titlepage}
\pagestyle{plain}
\setcounter{page}2
\baselineskip=17pt

\paragraph{1.}

The decays $\eta \to 3\pi$ are forbidden in the limit of isospin
conservation. Neglecting the small electromagnetic corrections
\cite{elmcorr}, the amplitudes are proportional to the isospin 
breaking mass difference $m_{u}-m_{d}$. The leading-order amplitude 
in the low-energy expansion of ${\cal O}(p^{2})$ \cite{osborne} is 
known to receive large higher-order corrections, both at 
${\cal O}(p^{4})$ \cite{GL85c} and beyond \cite{KWW96,AL96}.

The radiative decay $\eta \to \pi^{+}\pi^{-}\pi^{0}\gamma $ is in principle
an interesting channel. At lowest order $p^2$, the amplitude is pure
bremsstrahlung. At next-to-leading order an additional contribution appears
(direct emission) that is non-vanishing even in the isospin limit. Therefore,
the direct emission amplitude carries in principle new information that is
not accessible in $\eta \to 3\pi$ decays. The notion of a direct emission
amplitude is not unique except that it starts at ${\cal O}(k)$ where $k$
is the photon momentum. For instance, the so-called quadratic slope
parameters of the non-radiative amplitude arising at ${\cal O}(p^{4})$
also generate a radiative amplitude of ${\cal O}(k)$ that one may combine
with the bremsstrahlung amplitude because it is also completely fixed by the
non-radiative process. We have recently shown \cite{DEIN96} that one can
define a generalized bremsstrahlung (GB) amplitude for a generic radiative
four-meson process that includes the effects of all local terms of
${\cal O}(p^4)$ contributing to the non-radiative transition.

The main advantages of the GB amplitude are:

\begin{itemize}
\item  Since all local contributions to the non-radiative amplitude of $%
{\cal O}(p^{4})$ are included, the uncertainties in the corresponding
low-energy constants do not propagate to the direct emission amplitude
(defined here as the difference between the total and the GB amplitudes).

\item  If there are substantial higher-order contributions beyond
${\cal O}(p^{4})$ in the non-radiative amplitude they can be included in the GB
amplitude by using the experimentally measured non-radiative amplitude. For $%
\eta \to \pi ^{+}\pi ^{-}\pi ^{0}\gamma $, this is especially welcome
because the unitarity corrections \cite{KWW96,AL96} modify both rate and
slope parameters of $\eta \to 3\pi $ substantially. Using the experimental
values in the GB amplitude allows for a much more accurate determination of
the total amplitude.
\end{itemize}

The purpose of this letter is to calculate both GB and direct emission
amplitudes for $\eta \to \pi^{+}\pi^{-}\pi^{0}\gamma $ along the same lines
as for $K \to 3 \pi \gamma$ \cite{DEIN97}. We comment on the
differences between the GB and the usual bremsstrahlung amplitudes and we
discuss the relative importance of the main contributions to direct
emission: pion loops (isospin violating), kaon loops and vector meson
exchange (both isospin conserving). The suppression of the isospin
conserving component of direct emission is explained.

The channel under consideration has
already been studied in the framework of chiral perturbation theory by
Bramon et al. \cite{BGT96} where also references to the earlier literature
can be found. We will discuss the differences to Ref.~\cite{BGT96} as we go
along.

The upper limit $B(\eta \to \pi^{+}\pi^{-}\pi^{0}\gamma) < 
6 \times 10^{-4}$ quoted by the Particle Data Group \cite{PDG98}
refers to direct emission only \cite{Thaler}.
The experimental situation will improve considerably in the near
future. For instance, the KLOE experiment at 
the Frascati $\Phi$-factory \cite{DAFNE} should 
collect more than $10^{8}\;\eta$ per year.

\paragraph{2.}

To evaluate the bremsstrahlung contribution to $\eta
\rightarrow \pi^{+}\pi^{-}\pi^{0}\gamma $ we need to know the amplitude for $%
\eta(p_\eta) \rightarrow \pi^{+}(p_+)\pi^{-}(p_-) \pi^{0}(p_0)$ . Neglecting
electromagnetic corrections \cite{elmcorr}, the amplitude can be written in
the form \cite{GL85c} 
\begin{equation}
A(s,s_{\pm })=\frac{B(m_{u}-m_{d})}{3\sqrt{3} F_{\pi }^{2}}(1+3\frac{s-s_{0}%
}{M_{\eta }^{2}-M_{\pi }^{2}}) \left( 1+\delta(s,s_{\pm })\right)
\label{eta3piamp}
\end{equation}
where $B$ is a parameter of the lowest-order chiral Lagrangian \cite{GL85a}
related to the quark condensate and $F_{\pi}=92.4$ MeV is the pion decay
constant. The kinematical variables $s,s_{\pm },s_0$ are defined as 
\begin{equation}
s= (p_\eta-p_0)^2~, \qquad s_\pm = (p_\eta-p_\pm)^2~, \qquad s_0 = \frac{1}{3%
}(s + s_+ +s_-)~.  \label{eta3pikin}
\end{equation}
The function $\delta (s,s_{\pm })$ vanishes to lowest order $p^2$. At $%
{\cal O}(p^{4})$ it receives both loop and counterterm contributions \cite
{GL85c}. Higher-order effects due to $\pi\pi$ rescattering are important and
have been included in $\delta (s,s_{\pm })$ by way of dispersion relations 
\cite{KWW96,AL96}. These higher-order corrections increase the rate of $%
{\cal O}(p^{4})$ by some 25 $\div$ 30 $\%$ and must be included for a
reliable estimate of the bremsstrahlung amplitude.

Experimental results are conventionally expressed in terms of the Dalitz
variables $x,y$ defined as 
\begin{equation}
x = \frac{\sqrt{3}(s_{-}-s_{+})}{2M_{\eta }Q}~,\qquad y = \frac{3}{2M_{\eta
}Q}[(M_{\eta }-M_{\pi ^{0}})^{2}-s]-1~,  \label{eta3pidali}
\end{equation}
\[
Q=M_{\eta }-2M_{\pi ^{+}}-M_{\pi ^{0}}~. 
\]
Up to a normalization constant, the experimental Dalitz plot distribution is
fitted by a function of the form \cite{PDG98,charged} 
\begin{equation}
A(x,y)^2=A(0,0)^2(1+ay+by^{2}+cx^{2})~~  \label{eta3piexpt}
\end{equation}
where $A(x,y)$ corresponds to the decay amplitude (\ref{eta3piamp}).
Charge conjugation invariance forbids a term linear in $x$.

The present experimental and theoretical status of the parameters in (\ref
{eta3piexpt}) is summarized in Table~\ref{tab:slopes}. We do not need a
value for $A(0,0)$ since we always normalize our results to the
non-radiative decay. In this way, errors are substantially reduced. From
Table \ref{tab:slopes}, the importance of higher-order corrections is
evident also for the slope parameters. For the numerical calculation, we
will use the experimental values of $a,b$. Experiments have not been
sensitive enough to extract the parameter $c$ which is however relatively
stable with respect to chiral corrections (we will take $c=0.10$ for the
numerics).

\begin{table}[t]
\begin{center}
$
\begin{array}{|lr|c|c|c|}
\hline
\mbox{} &  & a & b & c \\ \hline
\mbox{}\mathrm{Experiment} & \cite{PDG98,charged} & -1.22\pm 0.07 & 0.22\pm
0.11 & - \\ \hline
\mbox{}\mathrm{Gasser\ and\ Leutwyler} \quad {\cal O}(p^{4}) & {\cite
{GL85c}} & -1.33 & 0.42 & 0.08 \\ \hline
\mbox{}\mathrm{Kambor\ et\ al.\ (solution\ a)} & {\cite{KWW96}} & -1.16 & 
0.24 & 0.09 \\ \hline
\mbox{}\mathrm{Kambor\ et\ al.\ (solution\ b)} & {\cite{KWW96}} & -1.16 & 
0.26 & 0.10 \\ \hline
\end{array}
$%
\end{center}
\caption{Experimental and theoretical values of the linear and quadratic
slopes of $\eta \rightarrow \pi ^{+}\pi ^{-}\pi ^{0}$ defined in Eq.~(\ref
{eta3piexpt}).}
\label{tab:slopes}
\end{table}

The kinematics of the decay $\eta(p_{\eta })\rightarrow
\pi^{+}(p_{+})\pi^{-}(p_{-})\pi^{0}(p_{0})\gamma(k)$ is specified by adding
the variables 
\begin{equation}
t_i = k \! \cdot \! p_i \qquad (i =\eta, +,-,0)  \label{kinema}
\end{equation}
with 
\[
t_\eta = t_+ +t_- +t_0~. 
\]
Any three of the $t_i$ together with $x$ and $y$ in (\ref{eta3pidali}) form
a set of independent variables.

With CP conserved, there is only an electric transition amplitude that we
write as 
\begin{equation}  \label{ampl}
A(\eta \rightarrow \pi^+ \pi^- \pi^0 \gamma) = e \varepsilon^\mu(k)
E_\mu
\end{equation}
with 
\[
k^\mu E_\mu = 0~. 
\]

Low's theorem \cite{Low58} relates the radiative amplitude to the
corresponding non-radiative amplitude and their first derivatives with
respect to the Dalitz variables up to ${\cal O}(k)$. For a general
four-body amplitude $A(s,t)$ with Mandelstam variables $s,t$, both $%
\displaystyle \frac{\partial A(s,t)}{\partial s}$ and $\displaystyle \frac{
\partial A(s,t)}{\partial t} $ contribute to the Low amplitude. Since there
are two neutral particles in our process we can choose variables and assign
particle labels such that only one of the derivatives enters. 
With $p_{1}=-p_{\eta }$, $%
p_{2}=p_{0}$, $p_{3}=p_{-}$ and $p_{4}=p_{+}$ in the notation of Ref.~\cite
{DEIN96}, Low's theorem reads 
\begin{equation}
\begin{array}{lll}
E_{\mathrm{Low}}^{\mu} & = & A(x,y)\left(\displaystyle \frac{%
p_{+}^{\mu }} {t_+}-\displaystyle \frac{p_{-}^{\mu }}{t_-}\right) \\ 
& - & \displaystyle \frac{\sqrt{3}}{M_{\eta }Q}\left[ p_{0}^{\mu }+p_{\eta
}^{\mu }-\frac{ p_{-}^{\mu }}{t_{-}}(t_{0}+t_{\eta })\right] \displaystyle 
\frac{\partial A(x,y)}{\partial x} + {\cal O}(k)
\end{array}
\label{brems}
\end{equation}
where
\begin{eqnarray} 
x &=& \frac{\sqrt{3}}{M_{\eta }Q}\left[p_\eta \czdot (p_+ - p_-)
+ t_- + t_0 \right] \nl
y &=& \frac{3}{2M_{\eta}Q}\left[(M_{\eta }-M_{\pi ^{0}})^{2}-
(p_+ + p_- + k)^2\right]-1
\end{eqnarray} 
from now on.

To lowest order $p^2$, the radiative amplitude is completely given by the
Low amplitude (\ref{brems}). In fact, since there is no $x$-dependence in
the $\eta \to 3 \pi$ amplitude of ${\cal O}(p^2)$ in (\ref{eta3piamp}),
only the non-derivative part in (\ref{brems}) contributes. Starting at $%
{\cal O}(p^4)$, an $x$-dependence is generated that produces the
quadratic slope term $c x^2$ in (\ref{eta3piexpt}).

However, one can do better than that. In order to account for all the local
parts of ${\cal O}(p^4)$ in the non-radiative amplitude that contribute
also to the radiative amplitude, a so-called generalized bremsstrahlung
amplitude can be introduced \cite{DEIN96}. One major advantage of using the
GB amplitude is that the remaining direct emission amplitude $E^\mu - E^\mu_{%
\mathrm{GB}}$ can only receive contributions from local terms of ${\cal O}
(p^4)$ that do not contribute to the non-radiative amplitude. For $\eta
\rightarrow \pi^{+}\pi^{-}\pi^{0}\gamma$, only the low-energy constant $L_9$ 
\cite{GL85a} could therefore appear in the direct emission amplitude.
However, the corresponding counterterm does not contribute to $\eta
\rightarrow \pi^{+}\pi^{-}\pi^{0}\gamma $ even for $m_u \ne m_d$. Thus, the
one-loop contribution to direct emission is necessarily finite.

The general formula for the GB amplitude of Ref.~\cite{DEIN96} simplifies in
the present case to 
\begin{equation}
\begin{array}{lll}
E_{\mathrm{GB}}^{\mu } & = & A(x,y)(\displaystyle \frac{p_{+}^{\mu}} {t_+}-%
\displaystyle \frac{p_{-}^{\mu }}{t_-}) \\ 
& - & \displaystyle \frac{\sqrt{3}}{M_{\eta }Q}\left[ p_{0}^{\mu }+p_{\eta
}^{\mu }-\frac{ p_{-}^{\mu }}{t_{-}}(t_{0}+t_{\eta })\right] \displaystyle 
\frac{\partial A(x,y)}{\partial x} \\ 
& + & \displaystyle \frac{3}{2M_{\eta }^{2}Q^{2}}\left\{ (t_{0}+t_{\eta
})\left[ p_{0}^{\mu }+p_{\eta }^{\mu }-\frac{p_{-}^{\mu }}{t_{-}}%
(t_{0}+t_{\eta })\right] -(t_{-}p_{+}^{\mu }-t_{+}p_{-}^{\mu })\right\} 
\displaystyle \frac{\partial ^{2}A(x,y)}{\partial x^{2}} \\ 
& - & \displaystyle \frac{3\sqrt{3}}{M_{\eta }^2 Q^2}\left(t_\eta p_{0}^{\mu
} -t_0 p_{\eta}^{\mu }\right)\displaystyle \frac{\partial^2 A(x,y)}{\partial
x \partial y} + {\cal O}(k)~.
\end{array}
\label{gbrems}
\end{equation}

If one uses the experimental amplitude as given by the Dalitz plot
distribution (\ref{eta3piexpt}) the last term in (\ref{gbrems}) involving $%
\displaystyle \frac{\partial^2 A(x,y)}{\partial x\partial y}$ will in fact
not contribute. As already announced, we use for the slope parameters the
experimental values \cite{PDG98,charged} $a=-1.22 \pm 0.07$, $b=0.22 \pm
0.11 $ and the theoretical prediction \cite{KWW96} $c=0.10$. The results for
the rate normalized to $\Gamma(\eta \to \pi^+ \pi^- \pi^0)$ are given in
Table \ref{tab:rateGB} for five bins in the photon energy $E_\gamma$ (in the 
$\eta$ rest frame).

\begin{table}[t]
\begin{center}
\begin{tabular}{|c|c|}
\hline
$E_\gamma$ (MeV) & $\Gamma_{\mathrm{GB}}(\eta \to \pi^+ \pi^-
\pi^0\gamma)/\Gamma(\eta \to \pi^{+}\pi^{-}\pi^{0})$ \\ \hline\hline
10--30 & $(2.30 \pm 0.04 ) \times 10^{-3}$ \\ \hline
30--50 & $(5.99 \pm 0.10 ) \times 10^{-4}$ \\ \hline
50--70 & $(1.85 \pm 0.04 ) \times 10^{-4}$ \\ \hline
70--90 & $(4.47 \pm 0.11 ) \times 10^{-5}$ \\ \hline
$>$~90 & $(5.00 \pm 0.14 ) \times 10^{-6}$ \\ \hline
\end{tabular}
\end{center}
\caption{Rates for $\Gamma(\eta \to \pi^+ \pi^- \pi^0\gamma)$ with the GB
amplitude (\ref{gbrems}) for different bins in the photon energy $E_\gamma$,
normalized to $\Gamma(\eta \to \pi^{+}\pi^{-}\pi^{0})$.}
\label{tab:rateGB}
\end{table}

The relative branching ratios for $E_\gamma \ge 10$ and 50 MeV, respectively
are 
\begin{equation}
\begin{array}{lcl}
B(\eta \to \pi^{+}\pi^{-}\pi^{0}\gamma; E_\gamma \ge 10 ~\mathrm{MeV})_{%
\mathrm{GB}} & = & (3.14 \pm 0.05)\times 10^{-3} B(\eta \to
\pi^{+}\pi^{-}\pi^{0}) \\ 
B(\eta \to \pi^{+}\pi^{-}\pi^{0}\gamma; E_\gamma \ge 50 ~\mathrm{MeV})_{%
\mathrm{GB}} & = & (2.35 \pm 0.05)\times 10^{-4} B(\eta \to
\pi^{+}\pi^{-}\pi^{0}) ~.
\end{array}
\label{BRGB}
\end{equation}
The errors given in both (\ref{BRGB}) and Table \ref{tab:rateGB} are due to
the experimental errors of the slope parameters $a,b$. These errors would of
course be much larger if we would not normalize to $\Gamma(\eta \to
\pi^{+}\pi^{-}\pi^{0})$. 

We can now make a first comparison with the work of Ref.~\cite{BGT96}.
Bramon et al. constructed a simple approximation to the Low amplitude (\ref
{brems}). They dropped the derivative term in (\ref{brems}) and took instead
the amplitude $A(x,y)$ of ${\cal O}(p^4)$ \cite{GL85c} at the center of
the Dalitz plot. In fact, they did not exactly use the amplitude of Ref.~%
\cite{GL85c} but increased the counterterm amplitude to account for the
discrepancy between the experimental rate and the predicted rate of $%
{\cal O}(p^4)$. With these assumptions, they obtain \cite{BGT96} 
\begin{equation}
\begin{array}{lcl}
B(\eta \to \pi^{+}\pi^{-}\pi^{0}\gamma; E_\gamma \ge 10 ~\mathrm{MeV})_{%
\mathrm{bremsstrahlung}} & = & 2.81 \times 10^{-3} B(\eta \to
\pi^{+}\pi^{-}\pi^{0}) \\ 
B(\eta \to \pi^{+}\pi^{-}\pi^{0}\gamma; E_\gamma \ge 50 ~\mathrm{MeV})_{%
\mathrm{bremsstrahlung}} & = & 1.85 \times 10^{-4} B(\eta \to
\pi^{+}\pi^{-}\pi^{0})
\end{array}
\label{rateBGT}
\end{equation}
In spite of the rather drastic approximations made, this prediction is quite
close to our result (\ref{BRGB}) that is based on the GB amplitude (\ref
{gbrems}) and on experimental input for the slope parameters. Of course, it
is difficult to assign an error to the prediction of Bramon et al. With the
errors given in (\ref{BRGB}) due to the experimental errors of the slope
parameters, our prediction for $B(\eta \to \pi^{+}\pi^{-}\pi^{0}\gamma;
E_\gamma \ge 10 ~\mathrm{MeV})$ is more than 6 standard deviations bigger
than that of Ref.~\cite{BGT96}. The discrepancy increases for larger values
of the cut in the photon energy.

Before attributing any significance to the predictions (\ref{BRGB}), we will
of course have to investigate the direct emission amplitude. Before doing
so, we compare the rates for the GB amplitude (\ref{gbrems}) with the ones
for the Low amplitude (\ref{brems}) in the same photon energy bins as
before. The results displayed in Table \ref{tab:GBLow} show that the
differences are rather small in all energy bins. This is due to the fact
that $E^\mu_{\mathrm{GB}} - E^\mu_{\mathrm{Low}}$ is only sensitive to the
quadratic slope parameter $c$ in (\ref{eta3piexpt}), numerically the
smallest of the three parameters. Nevertheless, the difference between GB
and Low is still bigger than the one-loop contribution to the direct
emission amplitude to which we now turn.

\begin{table}[t]
\begin{center}
\begin{tabular}{|c|c|}
\hline
$E_\gamma$ (MeV) & $(\Gamma_{\mathrm{GB}}- \Gamma_{\mathrm{Low}%
})/\Gamma_{\mathrm{GB}} $ \\ \hline\hline
10--30 & $2.0 \times 10^{-3}$ \\ \hline
30--50 & $8.7 \times 10^{-3}$ \\ \hline
50--70 & $1.9 \times 10^{-2}$ \\ \hline
70--90 & $3.3 \times 10^{-2}$ \\ \hline
$>$~90 & $4.9 \times 10^{-2}$ \\ \hline
\end{tabular}
\end{center}
\caption{Relative differences in the rates between GB and Low. Listed are
the quantities $\displaystyle\int\nolimits_{E_{\gamma}^{(1)}}^{E_{%
\gamma}^{(2)}} \left( \displaystyle \frac{ d\Gamma _{\mathrm{GB}}}{%
dE_{\gamma }} - \displaystyle \frac{d\Gamma _{\mathrm{Low}}}{dE_{\gamma }}
\right) dE_{\gamma} \left/ \displaystyle\int\nolimits_{E_{%
\gamma}^{(1)}}^{E_{\gamma}^{(2)}} \displaystyle \frac{d\Gamma _{\mathrm{GB}}%
}{dE_{\gamma }}dE_{\gamma } \right. $ for different bins in the photon
energy.}
\label{tab:GBLow}
\end{table}

\paragraph{3.}

\begin{figure}[tbp]
\centerline{\epsfig{file=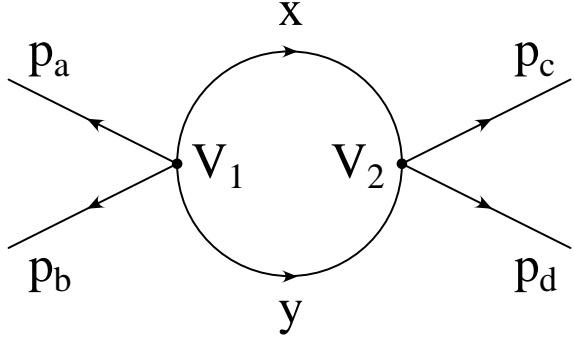,height=4cm}}
\caption{One-loop diagram for the general four--meson transition. For the
radiative amplitude, the photon must be appended to every charged meson line
and to every vertex with at least two charged fields. The vertices $V_1,V_2$
are defined in Eq.~({\ref{vertex}}).}
\label{fig:loop4}
\end{figure}

The full radiative amplitude is the sum of the GB amplitude (\ref{gbrems})
and of a direct emission amplitude $E^\mu_{\mathrm{DE}}$: 
\begin{equation}
E^{\mu }=E_{\mathrm{GB}}^{\mu }+E_{\mathrm{DE}}^\mu~.
\end{equation}
In this paragraph we calculate the direct emission amplitude of ${\cal O}
(p^{4})$. As shown in Ref.~\cite{DEIN96}, $E_{\mathrm{DE}}^\mu$ has the
following general structure at this order: 
\begin{equation}
E_{\mathrm{DE}}^{\mu } = E_{\mathrm{counterterm}}^{\mu }+ \sum_{\mathrm{loops%
}}(\Delta ^{\mu }+H^{\mu }) ~.  \label{EmuDE}
\end{equation}
As already mentioned, there is no counterterm contribution to direct
emission for $\eta \to \pi^+\pi^- \pi^0\gamma$. The (finite) loop
contribution is exclusively due to diagrams of the topology shown in Fig. 
\ref{fig:loop4} where a photon should be appended to all charged lines and
all vertices with at least two charged fields. 

The loop amplitude consists of a sum of two gauge invariant parts (for each
loop diagram) $\Delta^\mu$ and $H^\mu$. Referring to Ref.~\cite{DEIN96} for
details, we recall that both $\Delta^\mu$ and $H^\mu$ depend only on the
on-shell couplings of the vertices $V_1, V_2$ in Fig.~\ref{fig:loop4}. Those
vertices have the general form in momentum space 
\begin{equation}
\begin{array}{lcl}
V_1 & = & a_0 + a_1 p_a \! \cdot \! p_b + a_2 p_a \! \cdot \! x \\ 
&  & +\, a_3(x^2 - M_x^2) + a_4(y^2 - M_y^2) + a_5(p_a^2 - M_a^2) + a_6(p_b^2
- M_b^2) \\ 
V_2 & = & b_0 + b_1 p_c \! \cdot \! p_d + b_2 p_c \! \cdot \! x \\ 
&  & +\, b_3(x^2 - M_x^2) + b_4(y^2 - M_y^2) + b_5(p_c^2 - M_c^2) + b_6(p_d^2
- M_d^2)~.
\end{array}
\label{vertex}
\end{equation}
The relevant on-shell coefficients for the various diagrams are collected in
Table \ref{tab:coeff}. We have included the diagrams with two neutral
intermediate particles for completeness although they do not contribute to
either $\Delta^\mu$ or $H^\mu$ here. In general, $H^\mu$ is always zero in
this case but $\Delta^\mu$ may be non-zero depending on the assignment of
particle labels.\footnote{%
In fact, the loop contribution to $\Delta^\mu$ with two $\pi^0$ in the loop
was missed in the calculation of $K_L \to \pi^+ \pi^- \pi^0$ \cite{DEIN97}.
The change in the rate is numerically insignificant.}

The $\eta \to 3 \pi$ couplings vanish for $m_u=m_d$. In contrast
to Ref.~\cite{BGT96}, we keep the pion-loop contributions since they turn
out to be much bigger than the kaon loops which we calculate in
the isospin limit. The main contribution of direct emission arises in the
interference with the GB amplitude (\ref{gbrems}). The corresponding
contributions to the rate, separately for pions and kaons, are shown in
Table \ref{tab:piK}.

\begin{table}[t]
\caption{On-shell coefficients of the vertices $V_{1},V_{2}$ defined
in (\ref{vertex}) for the various loop diagrams of Fig. 1 in units of
$1/F^{2}$ and with $\overline{M_{1}^{2}}=(m_{d}-m_{u})B/ 
(\protect\sqrt{3}(M_{\eta}^{2}-M_{\pi }^{2}))$.}
\label{tab:coeff}
\[
\begin{array}{|c||c|c|c||c|c|c|}
\hline
\eta (-p_{b})\to \pi _{a}(p_{a}) &  &  &  &  &  &  \\ 
+\ M_{x}(x)M_{y}(y)\quad &  &  &  &  &  &  \\ 
\quad\ \to \pi _{c}(p_{c})\pi _{d}(p_{d}) & a_{0} & a_{1} & a_{2} & 
b_{0} & b_{1} & b_{2} \\ \hline
\eta \to \pi ^{0}~~+ &  &  &  &  &  &  \\ 
\pi ^{+}\pi ^{-}\to \pi ^{+}\pi ^{-} & -\overline{M_{1}^{2}}
(3M_{\eta}^{2}-M_{\pi }^{2})/3 & -2\overline{M_{1}^{2}} & 0 & 2M_{\pi }^{2}
& 2 & -2 \\ \hline
\eta \to \pi ^{0}~~+ &  &  &  &  &  &  \\ 
\pi ^{0}\pi ^{0}\to \pi ^{+}\pi ^{-} & -\overline{M_{1}^{2}}
(M_{\eta}^{2}-M_{\pi}^{2}) & 0 & 0 & M_{\pi }^{2} & 2 & 0 \\ \hline
\eta \to \pi ^{+}~+ &  &  &  &  &  &  \\ 
\pi ^{0}\pi ^{-}\to \pi ^{0}\pi ^{-} & 4\overline{M_{1}^{2}}M_{\pi }^{2}/3 & 
2\overline{M_{1}^{2}} & 2\overline{M_{1}^{2}} & M_{\pi }^{2} & 0 & -2 \\ 
\hline
\eta \to \pi ^{-}~+ &  &  &  &  &  &  \\ 
\pi ^{0}\pi ^{+}\to \pi ^{0}\pi ^{+} & 4\overline{M_{1}^{2}}M_{\pi }^{2}/3 & 
2\overline{M_{1}^{2}} & 2\overline{M_{1}^{2}} & M_{\pi }^{2} & 0 & -2 \\ 
\hline
\eta \to \pi ^{0}~+ &  &  &  &  &  &  \\ 
K^{-}K^{+}\to \pi ^{+}\pi ^{-} & M_{\pi }^{2}/(2\sqrt{3}) & \sqrt{3}/2 & 0 & 
0 & 0 & 1 \\ \hline
\eta \to \pi ^{0}~+ &  &  &  &  &  &  \\ 
K^{0}\overline{K^{0}}\to \pi ^{+}\pi ^{-} & - M_{\pi }^{2}/(2\sqrt{3}) & -%
\sqrt{3}/2 & 0 & 0 & 0 & 1 \\ \hline
\eta \to \pi ^{+}~+ &  &  &  &  &  &  \\ 
K^{0}K^{-}\to \pi ^{0}\pi ^{-} & M_{\pi }^{2}/\sqrt{6} & \sqrt{\frac{3}{2}}
& 0 & -M_{\pi }^{2}/\sqrt{2} & -1/\sqrt{2} & \sqrt{2} \\ \hline
\eta \to \pi ^{-}~+ &  &  &  &  &  &  \\ 
\overline{K^{0}}K^{+}\to \pi ^{0}\pi ^{+} & M_{\pi }^{2}/\sqrt{6} & \sqrt{%
\frac{3}{2}} & 0 & -M_{\pi }^{2}/\sqrt{2} & -1/\sqrt{2} & \sqrt{2} \\ \hline
\end{array}
\]
\end{table}

\begin{table}[ht]
\begin{center}
\begin{tabular}{|c|c|c|}
\hline
$E_\gamma$  (MeV) & $(\Gamma_{\mathrm{GB+DE}}- \Gamma_{\mathrm{GB}
})/\Gamma_{\mathrm{GB}}\quad$ (pions) &
$(\Gamma_{\mathrm{GB+DE}}-\Gamma_{\mathrm{%
GB}})/\Gamma_{\mathrm{GB}}\quad$ (kaons) \\ \hline\hline
10--30 & $-1.4 \times 10^{-4}$ & $0.7 \times 10^{-5}$ \\ \hline
30--50 & $2.4 \times 10^{-4}$ & $2.3 \times 10^{-5}$ \\ \hline
50--70 & $2.6 \times 10^{-3}$ & $5.7 \times 10^{-5}$ \\ \hline
70--90 & $9.4 \times 10^{-3}$ & $6.9 \times 10^{-5}$ \\ \hline
$>$~90 & $3.4 \times 10^{-2}$ & $5.8 \times 10^{-5}$ \\ \hline
\end{tabular}
\end{center}
\caption{Relative rate differences for the interference between GB and the
one-loop contributions to direct emission. The notation is analogous to
Table \ref{tab:GBLow}.}
\label{tab:piK}
\end{table}

The amplitude of ${\cal O}(p^4)$ is completely dominated by the pion
loops. We will explain the suppression of kaon loops
after the discussion of vector meson exchange.
Nevertheless, the residual pion-loop contribution in the direct emission 
amplitude is quite small for almost all photon energies. Integrating the
differential rate over the photon energy for $E_\gamma \ge 10$ MeV produces
a correction to the branching ratio that is smaller than the error given in 
(\ref{BRGB}) for the GB contribution only. It is even smaller than the
difference between the rates for GB and Low amplitudes 
(cf. Table \ref{tab:GBLow}). The relative size of the loop 
amplitude increases with $E_\gamma^{\mathrm{min}}$ at the expense of 
decreasing rates. 

For the loop contributions to direct emission we only agree with Ref.~\cite
{BGT96} to the extent that they are small. Bramon et al. did not include the
dominant pion loops and they did not calculate the interference with
the bremsstrahlung amplitude. Taking the kaon-loop amplitude by itself 
leads of course to a tiny rate that is completely negligible \cite{BGT96} in
comparison with the interference between the GB and the pion-loop
amplitudes.

\paragraph{4.}

Since there is no counterterm contribution to direct emission at 
${\cal O}(p^{4})$ resonance exchange can only enter at ${\cal O}(p^{6})$. 
Starting from the list of ${\cal O}(p^{3})$ vector and 
axial-vector couplings given in \cite{EGLPR89}, we have 
scanned all possible contractions of the resonance fields. 
In the isospin limit ($m_u=m_d$), the only 
surviving $\eta \to \pi^{+}\pi^{-}\pi^{0}\gamma$
amplitude of this type is generated 
through the product of the following vector operators: 
\begin{equation}
\cL_{V}\,=h_{V}\,\varepsilon _{\mu \nu \rho \sigma }\,\langle
\,V^{\mu }\,\{\,u^{\nu }\,,\,f_{+}^{\rho \sigma }\,\}\,\rangle \,+\,i\theta
_{V}\,\varepsilon _{\mu \nu \rho \sigma }\,\langle \,V^{\mu }\,u^{\nu
}\,u^{\rho }\,u^{\sigma }\,\rangle~,   \label{eq:vmd}
\end{equation}
where we have adopted the notation of \cite{Prades94} 
for the coupling constants. For the other combinations 
of resonance terms, either the $SU(2)$-singlet nature of the 
$\eta$ field or the resulting minimal number of pseudoscalar fields 
implies a vanishing contribution to the 
$\eta \to \pi^{+}\pi^{-}\pi^{0}\gamma$
amplitude.

The couplings in $\cL_{V}$ can in principle be determined from the
phenomenology of vector meson decays. The decay rate for $\omega \rightarrow
\pi^{0}\gamma $ \cite{PDG98} fixes $|h_{V}|=0.037$. For the second coupling 
$\theta _{V}$, one has to rely on models for the time being. Hidden symmetry
predicts $\theta _{V}=2h_{V}$ \cite{bando}, which is compatible with 
the value $\theta _{V}=0.050$ deduced from the  ENJL model \cite{Prades94}.
We shall also assume that the field $V^{\mu }$ in
(\ref{eq:vmd}) describes a nonet of vector mesons.

Integrating out the vector mesons in the Lagrangian (\ref{eq:vmd}), one
obtains the following effective Lagrangian of ${\cal O}(p^{6})$ for the
direct emission in $\eta \to \pi ^{+}\pi ^{-}\pi ^{0}\gamma$: 
\begin{equation}
\cL_{VMD}^{6}=\displaystyle \frac{64ih_{V}\theta _{V}F^{\mu \nu
}\partial^{\rho } \eta }{3\sqrt{3}M_{V}^{2}F_{\pi }^{4}}\left[ \partial
_{\rho }\pi ^{0} (\partial_{\nu }\pi ^{+}\partial _{\mu }\pi ^{-})+\partial
_{\mu } \pi ^{0}(\partial_{\rho }\pi ^{+}\partial _{\nu }\pi ^{-}-\partial_
{\rho }\pi ^{-}\partial_{\nu }\pi ^{+})\right]~.  \label{VMD6}
\end{equation}
This Lagrangian gives rise to the decay amplitude 
\begin{equation}
E_{\mathrm{DE,VMD}}^\mu = \displaystyle \frac{64 h_{V}\theta _{V}} {3\sqrt{3%
}M_{V}^{2}F_{\pi }^{4}}\left[p_\eta\cdot p_0 g^\mu_{+-} +p_\eta\cdot
p_+ g^\mu_{-0}+p_\eta\cdot p_-g^\mu_{0+}\right]
\label{VMDg}
\end{equation}
\[
g^\mu_{ij}=t_i p_j^\mu - t_j p_i^\mu 
\]
which differs from formula (19) in \cite{BGT96}.

For $h_{V}=0.037$ and $0.050 \leq \theta _{V} \leq 0.075$,
we find that this amplitude provides a contribution to direct 
emission that is smaller than the pion-loop amplitude, especially for
large photon energies. 
Actually each of the three separate gauge-invariant terms in
(\ref{VMDg}) generates a contribution which is of the same order 
or even larger than the one from the pion loops. However, there is a 
strong destructive interference among the three terms which leads to the
small results reported in Table \ref{tab:VMDtotal}. In the same Table 
we also show the total direct emission, obtained by summing loop and 
vector meson contributions. 

\begin{table}[t]
\begin{center}
\begin{tabular}{|c|c|c|}
\hline
$E_\gamma$  (MeV) & $(\Gamma_{\mathrm{GB+DE}}- \Gamma_{\mathrm{GB}%
})/\Gamma_{\mathrm{GB}}\quad$ (VMD) & $(\Gamma_{\mathrm{GB+DE}}-\Gamma_{\mathrm{GB%
}})/\Gamma_{\mathrm{GB}}\quad$ (total) \\ \hline\hline
10--30 & $0.6 \times 10^{-4}$ & $-0.7 \times 10^{-4}$ \\ \hline
30--50 & $3.2 \times 10^{-4}$ & $5.8 \times 10^{-4}$ \\ \hline
50--70 & $7.3 \times 10^{-4}$ & $3.4 \times 10^{-3}$ \\ \hline
70--90 & $1.1 \times 10^{-3}$ & $1.0 \times 10^{-2}$ \\ \hline
$>$~90 & $1.0 \times 10^{-3}$ & $3.5 \times 10^{-2}$ \\ \hline
\end{tabular}
\end{center}
\caption{Relative rate differences for the interference between GB and
direct emission: VMD (for $h_{V}\theta _{V}=2.8\times 10^{-3}$) 
and total direct emission. The notation is analogous
to Table \ref{tab:GBLow}.}
\label{tab:VMDtotal}
\end{table}

\paragraph{5.}

It is remarkable that the isospin conserving part of direct emission
(both kaon loops and vector meson exchange) is smaller than
the isospin violating component due to pion loops. In order to
understand this suppression, we write the decay amplitude in the form
\begin{equation}
E^\mu = A_{+-}(p_+,p_-,p_0,k) g^\mu_{+-} +
 A_{-0}(p_+,p_-,p_0,k) g^\mu_{-0} + A_{0+}(p_+,p_-,p_0,k) g^\mu_{0+}~. 
\label{Egeneral}
\end{equation}
Since only two of the $g_{ij}^\mu$ are linearly independent this
decomposition is of course not unique but it will be useful in the
limit $m_u=m_d$.

The first observation is that the amplitude $E^\mu$ vanishes when the
three pion momenta are equal. In the $\eta$ rest frame, this
configuration corresponds to maximal photon energy. Therefore, gauge
invariance alone implies that the complete amplitude and in particular
the direct emission part is small in a region where direct emission
has any chance at all against (generalized) bremsstrahlung.

Let us now consider the isospin limit $m_u=m_d$ where only (part of)
the direct emission amplitude survives. In this case, isospin
violation can only come from the electromagnetic field which is the 
sum of $I=0$ and $1$ spurions. Since $G =-1$ for
$\eta\pi^+\pi^-\pi^0$ only the $G=-1$, $I=0$ part of the photon
contributes. Therefore, the three pions must be in an isosinglet
combination and any two pions are in an $I=1$ state. This implies that 
the amplitude (\ref{Egeneral}) can be written in terms of a single
invariant function $A_{+-}$ with
\begin{eqnarray} 
A_{-0}(p_+,p_-,p_0,k) &=& A_{+-}(p_-,p_0,p_+,k) \nl
A_{0+}(p_+,p_-,p_0,k) &=& A_{+-}(p_0,p_+,p_-,k) \nl
A_{+-}(p_-,p_+,p_0,k) &=& A_{+-}(p_+,p_-,p_0,k) ~.
\end{eqnarray}
Both the $V-$exchange amplitude (\ref{VMDg}) and the kaon-loop
amplitude satisfy these conditions.

Expressing, e.g., $g_{-0}^\mu$ in terms of $g_{+-}^\mu, g_{0+}^\mu$,
the amplitude (\ref{Egeneral}) can be written as
\begin{eqnarray} 
E^\mu\mid_{m_u=m_d} &=&
~~~ \left[A_{+-}(p_+,p_-,p_0,k)-\displaystyle\frac{t_0}{t_+}
A_{+-}(p_0,p_-,p_+,k)\right] g^\mu_{+-} \nl
& & + \left[A_{+-}(p_0,p_+,p_-,k)
-\displaystyle\frac{t_-}{t_+}A_{+-}(p_0,p_-,p_+,k)\right] g^\mu_{0+} 
\label{Eud}
\end{eqnarray}
in the limit $m_u=m_d$. 
Therefore, the amplitude is doubly suppressed for large photon
energies: both the $g_{ij}^\mu$ and the two invariant amplitudes in
(\ref{Eud}) vanish in the symmetric configuration with $p_+=p_-=p_0$. 
In general, this is not the case for the explicitly isospin violating 
contributions proportional to $m_u-m_d$ as can be seen from 
Eqs.~(\ref{brems}), (\ref{gbrems}). 

The suppression of kaon-loop and vector meson exchange amplitudes
is therefore a general feature of the amplitude in the limit 
$m_u=m_d$, independently of the chiral order. The direct emission
amplitude is strongly dominated by the pion loops. Since
this residual pion-loop contribution is itself small compared to the 
dominant GB amplitude, the theoretical uncertainty of the total direct
emission amplitude is also small and certainly negligible in
comparison with the present experimental errors entering the GB 
amplitude.

\paragraph{6.}

Our main results can be summarized as follows:

\begin{enumerate}
\item[i.]  The concept of generalized bremsstrahlung is very efficient in
avoiding the propagation of uncertainties in the non-radiative decays to the
direct emission amplitudes. In the case at hand, we have shown that the
numerically important final state interactions in $\eta \to \pi ^{+}\pi ^{-}%
\pi ^{0}$ \cite{KWW96,AL96} are easily incorporated in the GB amplitude.
This allows for a very precise prediction of the radiative decay rate
normalized to the non-radiative transition:
\begin{equation}
B(\eta \to \pi^{+}\pi^{-}\pi^{0}\gamma; E_\gamma \ge 10 ~\mathrm{MeV}) 
= (3.14 \pm 0.05)\times 10^{-3} B(\eta \to \pi^{+}\pi^{-}\pi^{0}) ~.
\end{equation}

\item[ii.]  The $\pi \pi $ loops dominate the direct emission
amplitude even though they are isospin suppressed. Nevertheless,
the one-loop amplitude is negligible compared to GB
for most of the photon energy range.

\item[iii.]  Isospin conserving contributions to direct emission such
as kaon loops or vector meson exchange are suppressed, especially for
large photon energies. Since this is a general feature to all orders in
the chiral expansion we expect it to be very difficult if not
impossible to observe any direct emission 
effect in $\eta \to \pi^{+}\pi^{-}\pi^{0}\gamma$ even with the 
anticipated yield of $10^{8}\;\eta $ per year \cite{DAFNE}. 
\end{enumerate}

\section*{Acknowledgments}

We thank A. Bramon for discussions.
G.D. and G.I. acknowledge the hospitality of the 
Institute of Nuclear Theory (INT) at the University of Washington, 
where part of this work has been done.

\end{document}